\documentclass{article}
\usepackage[nonatbib, final]{nips_2017}

\usepackage[utf8]{inputenc} 
\usepackage[T1]{fontenc}    
\usepackage[colorlinks = true, urlcolor  =blue]{hyperref}
\usepackage{url}            
\usepackage{booktabs}       
\usepackage{amsfonts}       
\usepackage{nicefrac}       
\usepackage{microtype}      
\usepackage{float}
\usepackage{graphicx}
\usepackage{caption}
\usepackage{subcaption}
\usepackage[]{algorithm2e}
\usepackage{self}
\usepackage{times}
\usepackage{hyperref}
\usepackage{url}
\usepackage{amsmath}
\usepackage{amsfonts}
\usepackage{amssymb}
\usepackage{graphicx}
\usepackage{multirow}
\usepackage{booktabs}
\usepackage{subcaption}
\newcommand\blfootnote[1]{%
	\begingroup
	\renewcommand\thefootnote{}\footnote{#1}%
	\addtocounter{footnote}{-1}%
	\endgroup
}
\graphicspath{{img/}}
\title{Understanding and predicting travel time with spatio-temporal features of network traffic flow, weather and incidents}

\author{
	Shuguan Yang\\
	Dept of Civil and Environmental Engineering\\
	Carnegie Mellon University\\
	Pittsburgh, PA 15213 \\
	\texttt{shuguany@cmu.edu} \\	
	\And
	Sean Qian\\
	Dept of Civil and Environmental Engineering\\
	Heinz College\\
	Carnegie Mellon University\\
	Pittsburgh, PA 15213 \\
	\texttt{seanqian@cmu.edu}
}

\begin{document}

\maketitle
\begin{abstract}
	Travel time on a route varies substantially by time of day and from day to day. It is critical to understand to what extent this variation is correlated with various factors, such as weather, incidents, events or travel demand level in the context of dynamic networks. This helps a better decision making for infrastructure planning and real-time traffic operation. We propose a data-driven approach to understand and predict highway travel time using spatio-temporal features of those factors, all of which are acquired from multiple data sources. The prediction model holistically selects the most related features from a high-dimensional feature space by correlation analysis, principle component analysis and LASSO. We test and compare the performance of several regression models in predicting travel time 30 min in advance via two case studies: (1) a 6-mile highway corridor of I-270N in D.C. region, and (2) a 2.3-mile corridor of I-376E in Pittsburgh region. We found that some bottlenecks scattered in the network can imply congestion on those corridors at least 30 minutes in advance, including those on the alternative route to the corridors of study. In addition, real-time travel time is statistically related to incidents on some specific locations, morning/afternoon travel demand, visibility, precipitation, wind speed/gust and the weather type. All those spatio-temporal information together help improve prediction accuracy, comparing to using only speed data. In both case studies, random forest shows the most promise, reaching a root-mean-squared error of 16.6\% and 17.0\% respectively in afternoon peak hours for the entire year of 2014.\blfootnote{Published on \href{https://ieeexplore.ieee.org/document/8742547}{IEEE Intelligent Transportation Systems Magazine, Vol. 3, 221-234}}
\end{abstract}

\section{Introduction}
With the increase in travel demand, congestion has became an escalating issue nowadays impairing the efficiency of transportation systems. To mitigate this effect, substantial amount of efforts have been devoted into developing novel technologies and policies for traffic management in the last few decades. Central to the traffic management is the understanding of what factors affect travel time and/or traffic flow to what extent, and how to effectively predict travel time in real time. Moreover, the accuracy and reliability of the prediction usually plays an essential role in the deployment of those technologies and policies. Recent emerging sensing technologies bring in massive data from multiple data sources, which enable us to examine the travel time more closely. This research proposes a data-driven approach to holistically understand and predict highway travel time using massive data of traffic speeds, traffic counts, incidents, weather and events, all of which are acquired in real time from multiple data sources collected over the years. The prediction model selects the most related features from a high-dimensional feature space to better interpret the travel time that vary substantially both by time of day and by day to day.

Because of the rising demand for reliable prediction of traffic state/travel time, a number of methods have been proposed in the last two decades. Among them, machine-learning based methods, coupled with basic traffic flow mechanisms, are gaining popularities and becoming the mainstream in literature. Just to name a few representative studies, linear time series analysis has been widely recognized and used for traffic states forecasting, such as Linear regression\cite{zhang2003short},  Auto-Regressive Integrated Moving Average (ARIMA) \cite{pace1998spatiotemporal,kamarianakis2005space,kamarianakis2003forecasting} and its extensions, including KARIMA\cite{van1996combining} which uses Kohonen self-organizing map; Seasonal ARIMA \cite{williams1998urban}; Vector-ARMA and STARIMA \cite{kamarianakis2003forecasting}. Other examples include Kalman filtering \cite{guo2014adaptive}, non-parametric regression models \cite{smith2002comparison,rahmani2015non}, and Support Vector machines \cite{cong2016traffic,wu2004travel} for predicting travel time and flow. \cite{mitrovic2015low} exploited compressed sensing to reduce the complexity of road networks, then support vector regression (SVR) is used for predicting travel speed on links.  A Trajectory Reconstruction Model is used by \cite{ni2008trajectory} for travel time estimations. \cite{qi2014hidden} applied hidden Markov Model that incorporates traffic volume, lane occupancy, and traffic speed data. \cite{ramezani2012estimation} and \cite{yeon2008travel} also used Markov chains to predict travel time on arterial routes. A fuzzy logic model was adopted by \cite{vlahogianni2008temporal} to model the temporal evolutions of traffic flow. A hybrid empirical mode decomposition and ARIMA, or hybrid EMD-ARIMA model is developed by \cite{wang2016novel} for short-term freeway traffic speed prediction. In recent years, studies that use deep neural networks for traffic estimation start to emerge. For example, a restricted Boltzmann Machine(RBM)-based RNN model is used to forecast highway congestions \cite{ma2015large}. Stacked auto-encoder deep architecture is used by \cite{lv2015traffic} to predict traffic flows on major highways of California. In addition, a time Delay Neural Network (TDNN) model synthesized by Genetic Algorithm (GA) is proposed and used for short-term traffic flow prediction \cite{abdulhai2002short}. A stacked Restricted Boltzmann Machine(RBM) in combination with sigmoid regression is used for predicting short term traffic flow \cite{siripanpornchana2016travel}. Last but not least, a detailed review of short-term travel time prediction can be found in \cite{oh2015short} and \cite{wang2016novel}.


Despite of tremendous research on travel time/flow prediction, many studies focus on exploring temporal correlation at a single location, or spatio-temporal correlation on a small-scale network (such as a corridor network). In fact, traffic states of two distant road segments can be strongly correlated temporally. However, only a few studies have taken such spatial-temporal correlations into considerations when building prediction models for large-scale networks. For example, \cite{kamarianakis2003forecasting} considered spatial correlations as function of distance and degree of neighbors when applying multivariate autoregressive moving-average (ARIMA) model to the forecasting of traffic speed. Furthermore, \cite{kamarianakis2012real} discussed the extensions of time series prediction model by considering correlations among neighbors and the utilization of LASSO for model selection. \cite{zou2014space} introduced a space–time diurnal (ST-D) method in which link-wise travel time correlation with a time lag is incorporated. \cite{cai2016spatiotemporal} proposed a k-nearest neighbors algorithm (k-NN) model to forecast travel time up to one hour in advance. This model uses redefined inter-segments distances by incorporating the grade of connectivity between road segments, and considers spatial-temporal correlations and state matrices to identify traffic states. \cite{min2011real} proposed a modified multivariate spatial-temporal autoregressive (MSTAR) model by leveraging the distance and average speed of road networks to reduce the number of parameters.

%


%
Among literature, travel time/speed and traffic counts are the two most commonly used metrics to be predicted, oftentimes based on the features constructed by themselves in the prediction model. Information of other features relevant to traffic states, such as weather, road incidents, local events are rarely explored, which may also exhibit potential correlation and causality relations with congestion on road segments. Previous studies have shown that adverse weather conditions have detrimental effects on traffic congestion \cite{goodwin2002weather,sridhar2006relationship}. Moreover, different weather features can bring various levels of impacts on traffic delays \cite{maze2006whether}. In this study, we incorporate a complete set of weather features in the prediction model: temperature, dew point, visibility, weather type (rain/snow/fog etc), wind speed, wind guests, pressure, precipitation intensity and pavement condition (wet/dry). It has also been shown that travel time is sensitive to traffic incidents of various kinds \cite{cohen1999measurement,kwon2011decomposition}, including crashes, planned work zones and disabled vehicles. The actual impacts of incidents on the travel time of particular road segments depend on a number of features of the incidents, such as time, location, type, severity and the number of lanes closed. Thus, we will also take into consideration all those incidents features in our prediction model.

In addition, the exploration of spatio-temporal correlations of traffic states in literature are limited to simple metrics, such as the distance between road segments, the degree of connections and the number of time lags. They usually are determined exogenously and do not necessarily reveal the actual observations. In our approach, spatio-temporal features of the network and travel demand are more extensively explored from a variety of data sets by the data-driven approach, and thus the prediction model can adapt to the real-world traffic conditions in response to diverse roadway/demand disruptions. For example, in urban areas with a daily commuting pattern, time dependent origin-destination (O-D) travel demand in the morning peak is adopted as a feature when predicting travel time of the afternoon peak, since demand patterns in morning peak and afternoon are oftentimes correlated as will be shown later. Based on the O-D demand, alternative routes of the target road segments of prediction interest can be derived and incorporated into the prediction model to explore their spatio-temporal correlations.

The goal of this paper is two-fold: 1) analyze and interpret the spatio-temporal relation among highway congestion and various features such as weather, incidents, demand and travel speed in the context of dynamic networks, and 2) establish a reliable travel time prediction model for an arbitrary part of a large-scale network. The method incorporates features of both supply and demand including roadway network, travel demand, traffic speed, incidents, weather and local events, all of which are collected over several years.

Comparing to existing methods in literature, this paper makes the following contributions,
\begin{itemize}
	\item We consider a comprehensive list of data sets in the context of large-scale networks to extract features and explore their spatio-temporal relations with travel time. Those data sets include physical roadway networks, travel demand approximated by traffic counts, traffic speed, incidents, weather and local events, all of which are in high spatio-temporal resolutions and collected by time of day over the years. Existing studies usually focus on one single data set or a subset of them, on a small-scale network, and the spatial or temporal resolution is relatively coarse  \cite{sridhar2006relationship,kwon2011decomposition,cohen1999measurement}.
	\item The proposed prediction model is able to provide reliable results 30 minutes in advance. This is more advantageous than most short-term data-driven traffic prediction of 5-15 minutes ahead, such as ARIMA, Kalman filtering and decision tree \cite{chien2003dynamic,zhang2003short,kwon2000day,guin2006travel}, just to name a few.
	\item The two case studies are conducted for afternoon hours (2-6pm) on busy and unreliable corridors, when travel time varies the most significantly, both from day to day and within day. Many existing methods are applied to the entire day on mildly congested roads, which may partially alleviate the prediction challenge. In this sense, our model attempts to most effectively capture factors impacting traffic throughput and congestion evolution by analyzing the time of day period with the highest travel time variability.
	\item Performance of all prediction models, including a time-series model as a benchmark, are estimated through multi-fold cross validations in this paper, rather than separating training data set and testing data set in an ad-hoc manner. Cross validation results in a more robust model selection process and reliable estimators for model errors comparing to the conventional train/test validation in many studies.
	\item By exploring the spatio-temporal correlation among multi-source features, the travel time prediction model can be interpreted with findings and insights from real-world traffic operation.
\end{itemize}

The rest of this paper is organized as follow. The proposed method for data analytics and travel time prediction is introduced in Section 2. The method is then applied to the following two case studies: (1)A 6-mile highway corridor of I-270 Northbound near Washington, DC. (2)A 2.3-mile highway segment of I-376  around downtown Pittsburgh. Results and findings are presented in Section 3. Conclusions and future works are discussed in Section 4.

\section{Methodology}
\begin{figure}
	\centering
	\includegraphics[width=0.9\linewidth]{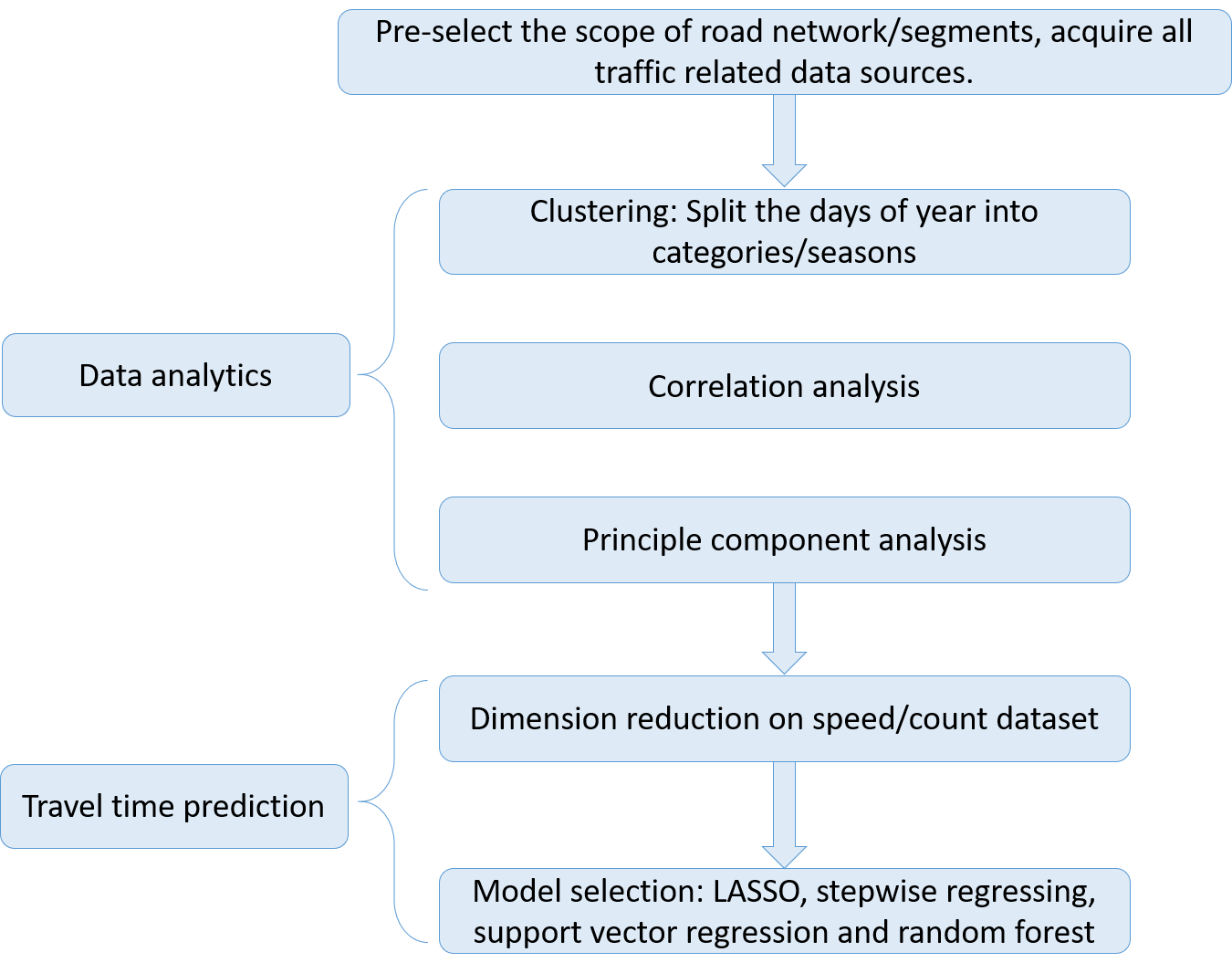}
	\caption{Procedure of data analytics and model selection}
	\label{data_proc}
\end{figure}

The proposed method has two main parts: data analytics and prediction model selection. The former aims to improving our understanding of the correlation among various features from multiple data sources and possible interpretation of congestion. The following methods are adopted: clustering, correlation analysis and principal component analysis. The latter part picks out a subset of features that are the most critical and robust in predicting travel time by incorporating the results from the analytics, prior knowledge of the network characteristics, and estimated recurrent travel demand. Finally, the best prediction model is selected out of several candidates, including LASSO (least absolute shrinkage and selection operator), stepwise regression, support vector regression and random forest. The overall procedure of the approach is shown in Fig \ref{data_proc}.


\subsection{Multiple data sources related to travel time}
\label{datasource}
As for short-term travel time prediction, two widely used data sources in literature are speed/travel-time and traffic counts, which are also known as direct indicators of real-time traffic states. To understand spatio-temporal correlations of speed and counts among road segments, this research incorporates features from all road segments in the region of interest when predicting travel time for each road segment, comparing to only the road segment itself or a few ones in its vicinity in most existing studies. The intuition is that many road segments, though being distant away from the road segment of interest, can be on its parallel routes to the route containing the segment of interest. Or more generally, the segment of interest may be impacted as a result of the ripple effect from some perturbations to some distant segments. Thus, examining only the segment itself and adjacent segments may overlook critical spatio-temporal correlation among distant road segments. The speeds and counts of all road segments, measured for each of the time-of-day intervals (such as 5 minutes) are used as features.

Apart from speeds and counts, other data sources have also been proven to effectively influence traffic congestion. Previous studies have shown that adverse weather conditions bring detrimental effects on traffic congestion \cite{goodwin2002weather,sridhar2006relationship}. Moreover, different weather features may bring in various levels of impacts on traffic delay \cite{maze2006whether}. In this study, we incorporate the following weather features in the model: temperature, dew point, visibility, weather type (rain/snow/fog etc), wind speed, wind guests, pressure, precipitation intensity and pavement condition (wet/dry). Travel time is also sensitive to traffic incidents of various kinds \cite{cohen1999measurement,kwon2011decomposition}, including crashes, planned work zones and disabled vehicles. The actual impacts on the travel time by an incident are dependent on a number of factors of the incident, such as time of day, location, type, severity and the number of lanes closed. In this study, multiple incident features for each road segment of study are extracted to encapsulate the location information of incidents to the segment, including several binary variables implying upstream or downstream, whether or not the incident is on its alternative routes, and whether or not the incident is along its opposite direction. Furthermore, the severity of incidents is incorporated into the set of incident features. The severity is considered with the following features: number of lane closed, number of motorists injured, and number of vehicles involved, all of which may be available from crash data. All those features of incidents are carefully examined in the correlation analysis and ultimately selected for the prediction model.

Last but not least, local events, such as sports games and festivals would alter the daily travel demand, and many influence congestion on relevant road segments. The actual impacts of an event on traffic are related to its date/time, location and scale. Due to the sparsity and heterogeneity nature of events, it may be infeasible to incorporate all those factors for each event without overfitting. In principle, events with potential correlations are manually selected. The event type, location and time are included in the set of features. For instance, one event feature can be a binary variable indicating whether a particular type of event takes place in the evening.



\subsection{Network characteristics and recurrent demand level}
The characteristics of a road network as well as the daily travel demand are considered when forming the initial set of features to be selected for relating traffic speed and counts.

First off, for any road segment of study, we look into those road segments that can potentially carry major recurrent traffic flow upstream or downstream. for the reason that their travel flow are most likely to have correlations with the segment of study. These segments can be extracted and selected by examining possible routes of the time-dependent origin-destination (OD) travel demand.

Second, correlations between segments on alternative routes for each OD pair are considered in our approach. The travel demand, and thus congestion level, on alternative routes are usually correlated, even if some segments are distant from each other. Correlations among such segments would otherwise not be learned when the degree of connections in a graph is used to capture correlations. Alternative routes are extracted from possible routes for each origin-destination pair.

Third, the day-to-day variations of daily commuting traffic are considered in this study. Travel demand during morning peaks can be highly correlated to demand during afternoon peak on the opposite direction of travel on the same day. When analyzing and predicting afternoon peak travel time, morning peak travel time on the other direction can be used as an indicator to approximate the demand level of daily commuting traffic.

To sum up, for the road segment of study, speed/counts features extracted from the following segments are included in the initial speed/counts feature set prior to model selection and prediction:
\begin{itemize}
	\item All upstream and downstream road segments that carry the same traffic flow along main routes with the segment of study, during the time period of prediction.
	\item All road segments on the alternative routes which are derived for all origin-destination pairs.
	\item When analyzing/predicting travel time of afternoon peaks, traffic counts on the opposite direction in morning peaks are included to approximate daily commute demand level.
\end{itemize}	

We further illustrate how we can approximate the daily demand level. Cumulative traffic counts, from early in the morning to immediately before morning congestion starts, may reveal the demand level. Therefore, the demand level can be approximated by traffic counts of those locations surrounding the segment of study, observed at the same of time of day period from day to day. The time period can start from as early as 4am when commuting starts, to the latest possible time of day throughout the entire year prior to the morning traffic breaks down, (e.g., 6:30am). A traffic break-down can be defined as the time when travel speed drops below a certain threshold, such as 40 miles/hour. Traffic counts during the congestion are limited by the flow capacity of the furthest downstream link, and thus cannot be directly used to estimate the daily demand level.

%
\subsection{Data analytics}
We apply several data analytics techniques to the multi-source traffic data to gain a better understanding of how features are related to congestion, and provide insights for building a reliable prediction model.
\subsubsection{Clustering}
\label{cluster_section}
In general, daily traffic pattern usually owns day-of-week effects and/or seasonal effects. For instance, winter or summer, monday, Tuesday through Thursday, and Friday, days within each of those category may exhibit similar patterns. Thus, we cluster days of year into several categories, then conduct data analytics and travel time prediction for each cluster independently.

W can first apply K-means clustering to separate days into several clusters. Then, depending on the distribution of days of week in each cluster, we determine whether or not it is suitable to aggregated data based on their days of week. For example, if most Mondays fall into one cluster and rarely appear in another cluster, we can infer that Mondays exhibit a unique traffic pattern from other days and should be analyzed separately. The objective of the K-means clustering on observations $(x_1, x_2, \dots, x_n)$ is to find $k$ clusters of sets $S$, which satisfies
	\begin{equation}
	\arg \min_S \sum_{i=1}^{k}\sum_{x \in S_i} || x-\mu_i||^2 = 	\arg \min_S \sum_{i=1}^{k}|S_i|\text{Var} S_i
	\label{kmeans}
	\end{equation}
	where $\mu_i$ is the mean of all observations in set $S_i$.

In another example, we would like days from the same month/seasons to be grouped together. In this case, we can apply hierarchical agglomerative clustering (HAC) \cite{zepeda2013hierarchical}, which aims to minimize the within-cluster sum of squared error, with additional constraints that observations in the same cluster must be in a connected graph. In other words, we can enforce that all of the days within the same cluster form a continuous range of dates.

In terms of the number of clusters, our method explores a range of options, and selects one based on the goodness of fitting as well as the dimension of the training set. In the case that the size of training set is large enough, e.g. daily observations of more than three years are available, we can split data into more clusters. For hierarchical agglomerative clustering, it is essential that each cluster contains a sufficient amount of days to avoid over-fitting.

\subsubsection{Correlation analysis}
To identify the relationship among different traffic features, correlation analysis is conducted first. The Pearson's product-moment coefficient is defined by,
	\begin{equation}
	\rho_{X,Y} = \text{corr}(X,Y) = \frac{\text{cov}(X,Y)}{\theta_X \theta_Y}=\frac{E[(X-\mu_X)(Y-\mu_Y)]}{\theta_X \theta_Y}
	\label{corr}
	\end{equation}

By calculating the correlation matrix, and conduct hypothesis testing on whether certain pairs of features are correlated, we analyze the data in the following ways:
\begin{itemize}
	\item Calculate and evaluate the correlation among speed/count features of all road segments. We also explore the relationship of congestion among different road segments under various time lags. This helps us analyze how congestion propagates spatially and temporally.
	\item Test and analyze the correlation between incidents, weather features and the travel time on the segment of study. This helps us determine if these factors are correlated with congestion, and to what extent hazardous conditions, such as crash and wet surface, impact the segment of study.
	\item Correlation analysis results can also be used for feature selection. High correlation between features indicate redundancy in the feature set. In particular, if the two road segments exhibiting highly correlated speed/counts are adjacent to each other, we can either remove one of them, or replace both with their average in the feature set.
	\item As we approximate daily travel demand level using morning traffic counts, correlation analysis helps determine which road segments and time periods are the most critical in predicting afternoon-peak travel time. Apart from comparing the correlation coefficients and conducting hypothesis tests, we also plot the day-to-day morning traffic counts against  afternoon-peak counts, in order to infer if the morning counts are useful.
\end{itemize}

\subsubsection{Principal component analysis}
The selected features can be further explored by principal component analysis (PCA). PCA can break the entire set of features into several uncorrelated components via an orthogonal transformation. By conducting a PCA, the most important sources of variations from all the features can be found. PCA also allows us to compress the high-dimensional data by aggregating features into several critical dimensions.

In our method, a PCA can be conducted by first gathering all initial speed/counts features as well as the travel time of the segment of study, combining them with other features of incidents, weather and local events, and applying singular value decomposition \cite{abdi2010principal} to all features. Finally, we sort all principal components by their importance. Next, we can analyze the composition of the top few principal components to understand the main source of variations in the feature space. Also, by comparing the top principal components between different clusters of days, we are able to discover which features are the keys to distinguish clusters.

\subsection{Prediction model}
Based on the results of data analytics, a travel time prediction model can be established in the following steps.

\subsubsection{Dimension reduction (feature selection) from all the speed/counts features}
\label{221}
The number of features available from the aforementioned data sources are excessive comparing to the number of available data points. For instance, a segment of interest on I-270 northbound (to be shown later in the case study) has over 500 road segments of speed measurements in the regional network. When using five 5-min time lags, the prediction model has over 2,500 features from speed data only, all of which may have some degree of correlations with the travel time on I-270. There are around 260 workdays within a year that have afternoon peaks no longer than 5 hours. Predicting travel time with all those features, let alone features on weather/events/incidents, can be computational inefficient and more importantly, undertakes the risk of over-fitting. Hence, it is essential to reduce the dimension of the feature space before applying a prediction model.

The number of speed/counts features in the initial feature set is dependent on the complexity of road network and the number of time lags considered. For a particular segment of interest, a significant portion of those features are uncorrelated or redundant. Various regression models can be used to pick out those redundant features, such as rigid regression and LASSO. For LASSO, by tuning the $\lambda$ value in the formula below, we make trade off between the resulted dimension of features and the bias of the mean estimator.
\begin{equation}
\min_{\beta \in \mathbb{R}}\{\frac{1}{N}||y-X\beta||^2_2 + \lambda||\beta||_1\}
\label{lassoequ}
\end{equation}
Where $X, y, \beta$ are features, travel time of the segment of study, and the coefficients of features, respectively. For rigid regression, we can set a threshold for coefficients, below which the corresponding features are removed. We can achieve a similar flexibility by adjusting this threshold value. Although there is no strict rule on the amount of features to be retained in the final prediction model, the following factors can serve as metrics in determining a proper number: the size of the training data set, the complexity of the road network, and the expected minimum percentage of variation to be explained by the selected features (r-squared value). Notice that it is safe to leave slightly more features than it is necessary, since subsequence steps will further reduce the dimension of the entire feature space when weather/incidents/events are added.

While regression-based methods can pick out features that are linearly correlated with the travel time on the segment of study, those models may not reveal the actual possibly non-linear relationship. Moreover, data may be noisy and relatively sparse, it may not be reliable to rely solely on those data-driven methods. Thus, besides regression models, we also need to carefully review the selected features, and make modifications if necessary. For example, if there is no segment selected along the downstream/upstream of the segment of study, we should add some of them into the feature set manually, as they may exhibit non-linear relation with the target segment, or their relations are hidden by other highly correlated features chosen by regression. After this, all selected features will be used to create a non-linear regression model, such as a random forest.

\subsubsection{Model selection}
After pre-selecting the speed/counts features, we add weather/incidents/events features to construct a comprehensive prediction model. Correlation analysis is first conducted, to help select features that are highly correlated with the travel time of the target segment. Next, we apply a regression model to those selected features only. In this research, three models are adopted: LASSO linear regression, stepwise regression and random forest. Finally, we choose the one with the best prediction performance evaluated through cross validation as the final prediction model.

\section{Case studies}
To evaluate the performance of our method and explore in-depth insights from data analytics, we conduct two case studies. Details and results of the two case studies are presented in the next two subsections.

\subsection{I-270 Northbound}
The first case study is conducted on a 6-mile-long corridor of I-270 Northbound, between Montgomery Ave. and Quince Orchard Rd. in D.C. metropolitan region. The stretch of highway of interest is marked in purple in the left figure of Fig. \ref{DC_map}. The time period of study is 2PM-6PM on weekdays.
I-270 is a major corridor connecting D.C. metropolitan area with municipalities northwest of DC, such as Frederick and Gaithersburg. This segment is frequently congested during afternoon peak hours, and has substantial travel time variations from day to day and within days.

\subsubsection{Data sources and features}
As for the two case studies, travel time on road segments is measured in the manner of travel rate, which is the travel time over the entire stretch of study divided by their total length, namely the reciprocal of space mean speed. An initial features are constructed as follows:
\begin{enumerate}
	\item Speed: TMC (Traffic Message Channel) based speed data from INRIX is used, in all 369 TMC segments which covers all major highways and arterials of the area of study. Historical data are available in 5-minutes intervals. Those TMCs are shown in the right map of Fig \ref{DC_map} and listed below:
	\begin{itemize}
		\item Upstream/downstream of I-270, both Northbound and Southbound.(Orange on the map)
		\item Roadway network of northern D.C. area.(Blue on the map)
		\item MD 335 Northbound, as an alternative route of I-270 Northbound for afternoon peaks.(Red on the map)
		\item I-495 North, a major eastbound and westbound highway of northern D.C. area.(Green on the map)
	\end{itemize}
	We will predict travel time of the stretch of study in 30 minutes advance in the real time. 6 time lags are considered, from 60 minutes to 30 minutes in advance in 5-min intervals.
	
	\begin{figure*}[!t]
		\centering
		{\includegraphics[width=2.5in]{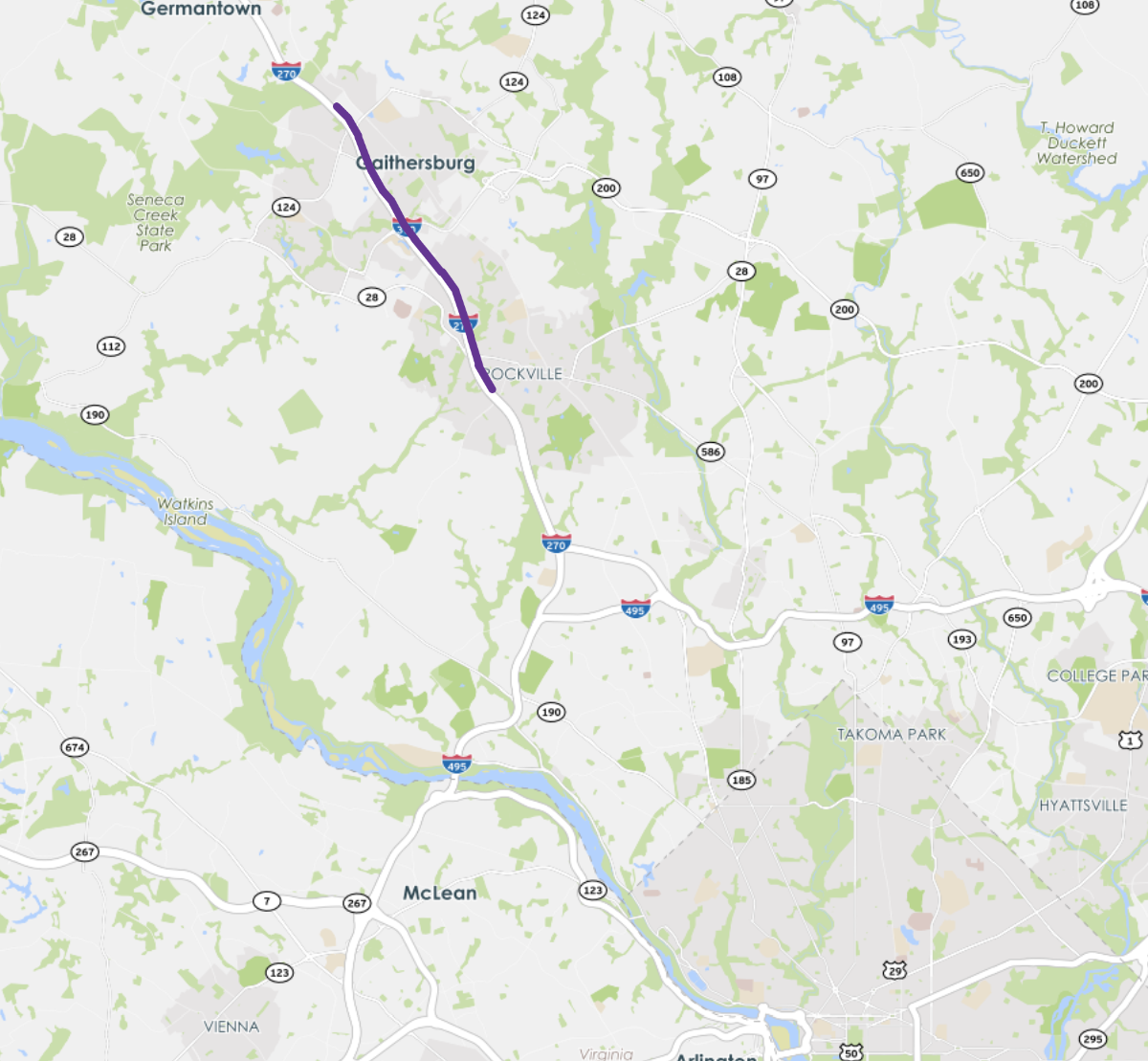}}
		\hfil
		{\includegraphics[width=2.5in]{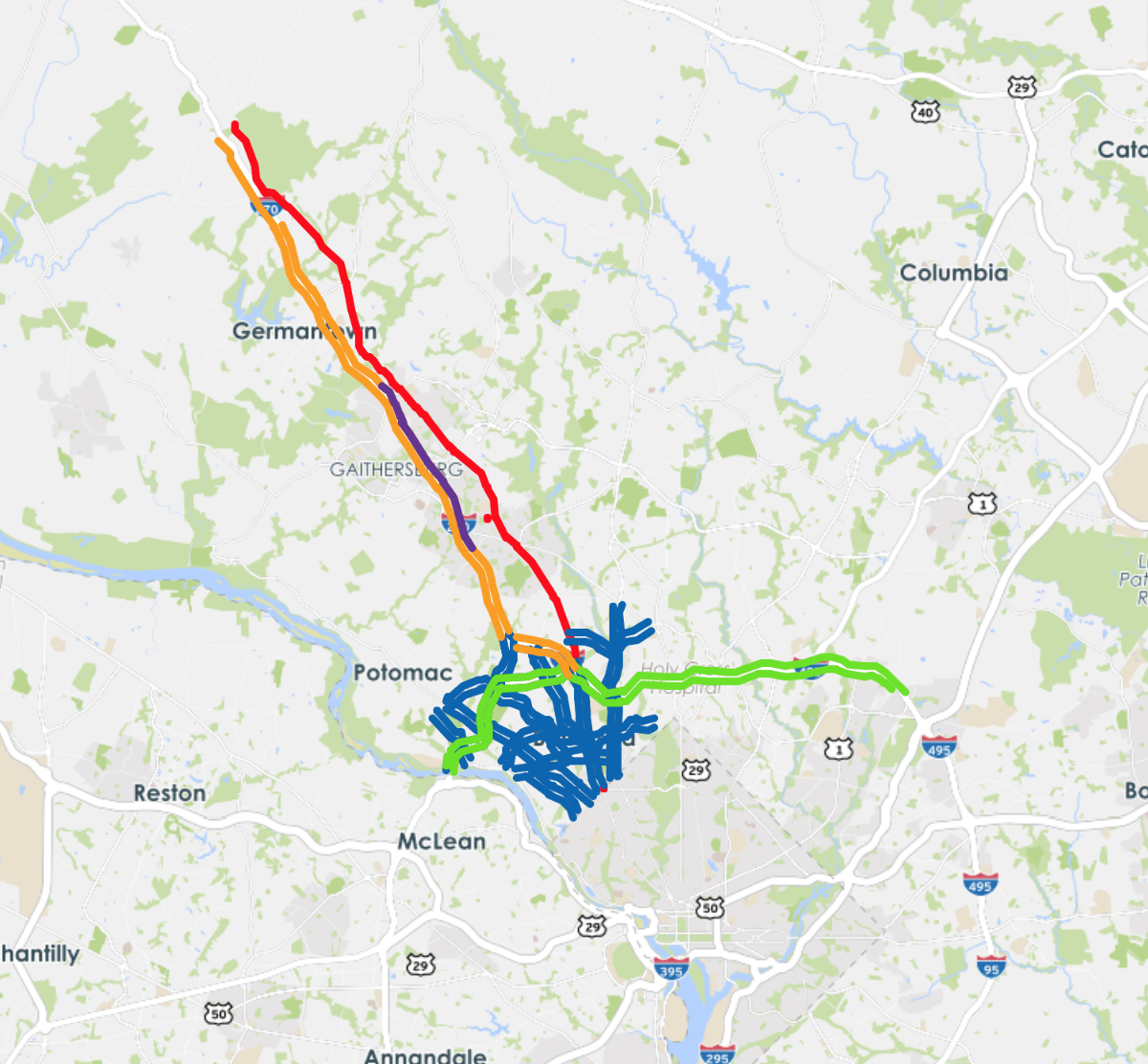}}
		\caption{Left: A 6-mile corridor of I-270 Northbound; Right: Speed data: 369 TMC segments}
		\label{DC_map}
	\end{figure*}
	
	\item Traffic counts: 5-min traffic flow counts from fixed location sensors on multiple locations of I-270 Northbound and Southbound are used. We use data from four of those sensors with the best data quality, two from each direction of this corridor. Morning and afternoon demand levels are estimated using the approximation method discussed in Section \ref{221}, namely demand level is approximated by cumulative counts in the early morning when congestion is not yet formed. In this case study, it is the aggregated counts on I-270 southbound, from 4AM to the earliest time of all days in 2017 when speed drops below 90\% of the free flow speed. Data from multiple sensors during this time period are summed up as the final value. The afternoon demand level is also calculated as the cumulative counts from 12PM to 3 PM.
	\label{inci_feature}
	\item Incidents: we use incidents data collected by the following state Department Of Transportation: Washington D.C., Maryland, and Virginia. Each incident is classified into one of the categories: crash, emergency road works, planned work zone and disabled vehicles. Each incident entry comes with start and end time, either provided in data or imputed. Note that we predict travel time/rate 30 min ahead, and thus, incidents reported within 30 minutes ahead at the time of prediction cannot be used as features. Based on the severity and geographical information of each incident, the following binary features are created:
	\begin{itemize}
		\item An incident on the upstream of the segments of study;
		\item A severe incident on the downstream of the segments of study;
		\item A non-severe incident on the downstream of the segments of study;
		\item An incident on the opposite direction (I-270 S).
		\item An incident on the downstream of the segments of study that is at least 3 miles away.
		\item An incident on the alternative route (MD 335 N);
		\item An incident in northern D.C. (the far upstream of the segments of study).
	\end{itemize}		
	In particular, severe incidents are defined as those with personal injuries reported. Upstream I-270 includes all road segments within 3 miles to the south end of the segments of study, while downstream I-270 includes those within 3 miles to the north end as well as the stretch of study itself. The reason for separating downstream and upstream is that downstream/on-site incidents usually reduce traffic speed on this stretch as a result of queues, while upstream incidents can reduce incoming flow rate to the stretch which may, in turn, result in an increase of traffic speed. Incidents on the opposite direction is defined on segments of I-270 Southbound within the same range of latitudes as the stretch of study (I-270 N). Finally, alternative routes are defined as those segments of MD 335 Northbound in the Rockville and Gaithersburg area.
	
	\item Weather: Hourly weather reports from Weatherunderground\footnote{https://www.wunderground.com/} is used in this case study. The following weather features are incorporated in the initial feature set: temperature (degrees Fahrenheit), wind chill temperature, precipitation intensity (inch/hour), precipitation type (Snow/Rain/None), Visibility (miles), wind speed (miles/hour), wind gust (miles/hour), pressure (millibar), Pavement Condition type (wet or dry). Also, we added categorical features of wind, visibility and precipitation intensity into the feature set. Those features are binary indicators whether the current condition is among the top 20\% extreme cases of the entire year of 2014.
	
	\item Local events: we incorporate the schedule of the following events: MLB (Washington DC nationals); NFL (Washington Redskins); NBA (Washington Wizards); NHL (Washington Capitals); DC cherry blossom festival. The event feature is binary and constant for the entire PM peak, indicating whether there is an ongoing/incoming event on that day.
\end{enumerate}

\subsubsection{Clustering}
We selecte in all 30 TMC segments from the speed feature set based on the following criteria: 1) The set of selected TMCs should be representative for the network. Thus, every major highway/arterial has at least one TMC selected; and 2) TMCs with a higher correlation with the travel time on the corridor of study are selected with high priority. We use speed measurements of three time points, 2:00PM, 4:00PM and 6:00PM for the 30 TMCs, in all 90 features for each weekday in 2013.

We test both K-means Clustering and hierarchical agglomerative clustering (HAC). From the results of K-means, we discover that day of week can not be properly separated by clustering as the composition of all cluster are a mixture of different days of week. This is probability due to the high variation nature of congestion in this corridor. With HAC, we discover that it is feasible to separate the whole year into two seasons:(1) 2013-02-21 to 2013-11-04; (2) 2013-01-01 to 2013-02-20 and 2013-11-05 to 2013-12-31. The two clusters can be explained as the non-winter pattern and the winter pattern. As a result, conduct data analytics and regression for each of the two clusters independently.

\subsubsection{Correlation analysis}
\begin{figure}
	\centering
	\includegraphics[width=1.05\linewidth]{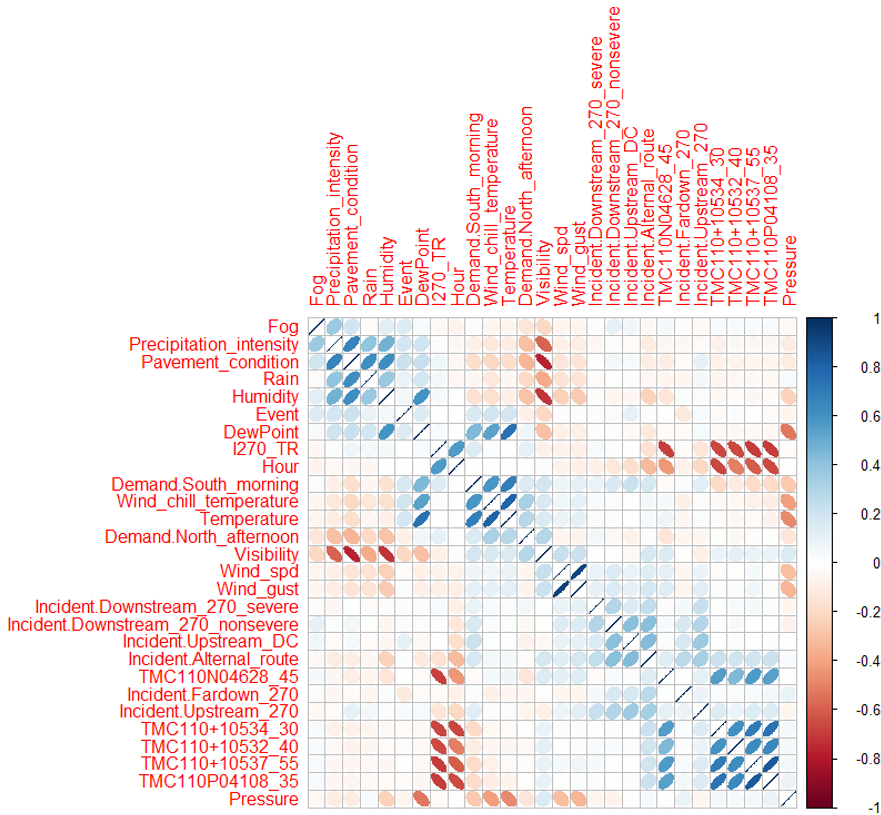}
	\caption{Correlation plot of selected features for winter season}
	\label{Corre_map}
\end{figure}

The correlation matrix of top five TMC features with the lowest p-values and a subset of non-TMC features are visualized in Fig. \ref{Corre_map}, due to the limit of space, not all features are visualized in the figure. Hypothesis tests (significant level 0.05) are also conducted for whether selected features are correlated with the travel rate on the corridor of interest. Findings are listed below:
\begin{itemize}
	\item Most TMC features are significant under the hypothesis tests, and the correlations between some TMC features and the travel rate(travel time) on target segments reach 0.7 in absolute values, much higher than all non-TMC features.
	\item Travel demand level in the morning are positively correlated with targeted travel rate, implying that morning travel demand can reveal afternoon congestion to some degree and can be used to predict travel time in afternoon peaks.
	\item Incidents on the upstream I-270 have a negative correlation with the targeted travel rate. In other words, when there is an incident upstream of I-270, downstream segments will experience less congestion than on a regular day.
	\item Presence of incidents on alternative route (MD 335) is also negatively correlated with the travel rate on the target corridor.
	\item Both severe and non-severe incidents on downstream I-270 are positively correlated with the targeted travel rate. The correlation coefficient of severe incidents is approximately 4 times as much as that of non-severe incidents. Overall, we see that the time, location and severity of incidents impact the congestion in very different ways.
	\item Among all weather features, wind speed, wind gust, visibility, precipitation intensity, rain, snow and pavement condition are significant under their hypothesis tests. The test on visibility owns the lowest p-value as 2.34e-09. This may reveal the causal effect of hazardous weather condition on congestion.
	\item Speed features are positively correlated with each other, including the speed on the corridor of study  (inverse of the travel rate).
	\item Rain and snow weather conditions have relatively high correlation with pavement condition, precipitation intensity and visibility, as expected.
	\item Travel rate has a positive correlation with the hour of day. More specifically, the probability of congestion increases as time progresses from 2:00PM to 6:00PM.			
\end{itemize}

\subsubsection{Principal component analysis}
\begin{figure*}[!t]
	\centering
	\includegraphics[width=5 in]{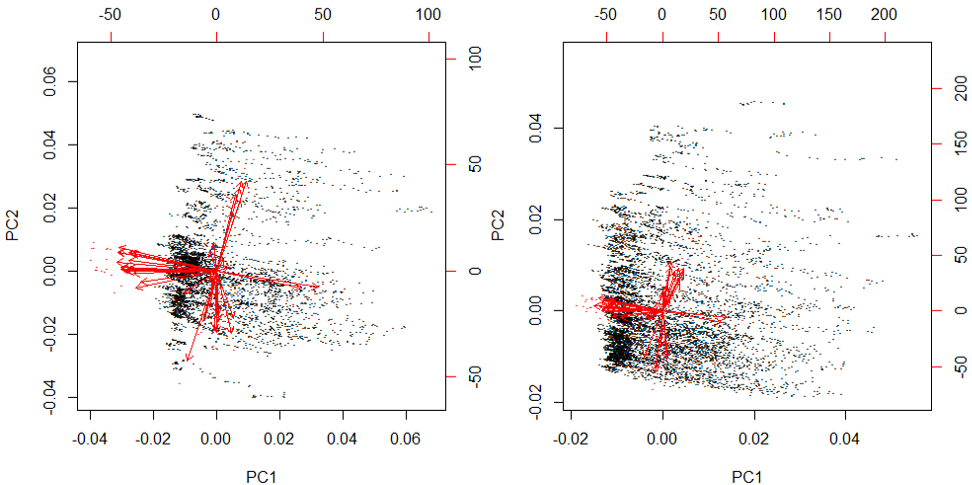}
	\caption{The first two principle components. Left: Non-winter cluster; Right: Winter cluster. Each black dot stands for a data entry. Red arrows are the loadings of all features}
	\label{PCA}
\end{figure*}

To find the sources of variation in the feature set, we conduct PCA. The PCA is conducted for the two seasons separately with the same set of features, including the five most correlated TMCs as well as features of counts, weather, incidents and events.  The first two principal components (PC), which can be interpreted as the two most important dimensions of the data are plotted in Fig \ref{PCA}. Each black dot in the plot is one data entry mapped to the orthogonal space of the two PCs. From the plots we see clear distinction between the two seasons, which indicates the existence of seasonal effects and further justifies the necessity for clustering.

In terms of the composition of the principal components, the first PC of the two clusters contains all five TMC-based speed features, and both account for around 35\% of the total variance. The second PC consists of morning and afternoon demand features, downstream incidents, and several weather features including precipitation type, visibility and pavement condition, accounting for another 10\% of the variance. The third PC is a mixture of incidents, weather and TMC-based speed, accounting for 7\% of the total variance. To conclude, TMC-based speed features are the most essential source of variance in the data, followed by demand level approximated by counts, downstream incidents and weather conditions.

\subsubsection{Dimension reduction (feature selection) in the TMC-based speed data set}
\label{select_DC}
We utilize LASSO to select a subset of TMC-based speed features in predicting travel rate/time. As discussed in section \ref{221}, by adjusting $\lambda$ in Equation \ref{lassoequ}, we obtain selected features and prediction results with different degrees of freedom. The degrees of freedom and corresponding r-squared values from LASSO, based on the winter cluster, is shown in Table \ref{T1}. Here, we also test the influence of prediction time lag to the model performance, and calculate the r-squared values under the same degree of freedom with 15min or 30min time lag, respectively. To predict travel time 30min ahead, we use speed features that are 30min, 35min, ... 55min ahead. Likewise, to predict travel teim 15min ahead, we use more speed features that are 15min, 20min, ... 55min ahead. 
\begin{table}[t!]
	\centering
	\renewcommand{\arraystretch}{1.2}
	\begin{tabular}{|l|l|l|}
		\hline
		& \multicolumn{2}{l|}{R-squared values} \\ \hline
		Degree of freedom & 30 min lag           & 15 min lag           \\ \hline
		0                 & 0.0000            & 0.0000            \\ \hline
		1                 & 0.2165            & 0.2165            \\ \hline
		2                 & 0.3532            & 0.3543            \\ \hline
		5                 & 0.4621            & 0.5358            \\ \hline
		7                 & 0.5422            & 0.6138            \\ \hline
		10                & 0.5971            & 0.5824            \\ \hline
		13                & 0.6328            & 0.7480            \\ \hline
		14                & 0.6561            & 0.6770            \\ \hline
		16                & 0.6709            & 0.7577            \\ \hline
		18                & 0.6808            & 0.7269            \\ \hline
		21                & 0.6873            & 0.7642            \\ \hline
		24                & 0.6920            & 0.7582            \\ \hline
		30                & 0.6951            & 0.7727            \\ \hline
		33                & 0.6973            & 0.7744            \\ \hline
		45                & 0.7011            & 0.7801            \\ \hline
		61                & 0.7050            & 0.7823            \\ \hline
		104          & 0.7116            & 0.7893            \\ \hline
		148           & 0.7186            & 0.7937            \\ \hline
		190           & 0.7246            & 0.7979            \\ \hline
		244           & 0.7301            & 0.8019            \\ \hline
	\end{tabular}
	\caption{Speed feature selection for winter season using LASSO }
	\label{T1}
\end{table}

Generally R-squared values increase as more speed features are selected. When predicting travel time 30min in advance, the marginal improvement in r-squared value starts to decline when the degree of freedom exceeds 18. Under the same degree of freedom, predicting travel time 15 min ahead are significantly more accurate than 30min ahead. Similar results are also observed for the non-winter cluster.

\begin{figure}
	\centering
	\includegraphics[width=4.5 in]{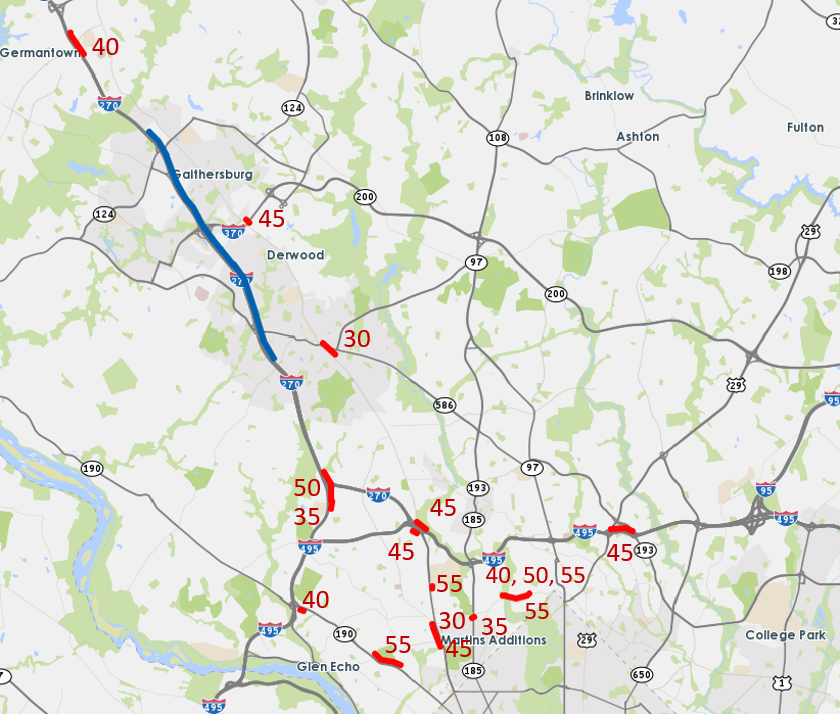}
	\caption{TMC segments selected for the winter season to predict travel time on I-270 northbound. Time lags in minutes are listed for each selected TMC.}
	\label{speed_DC}
\end{figure}

To the balance out the model's reliability (namely to avoid overfitting) and goodness of fit, we choose 18 speed features from 16 different TMC segments as a result of LASSO for the winter season. In Fig \ref{speed_DC}, the corridor of study is marked in blue, and those selected TMC segments are marked in red with time lags in minutes listed. For instance, 35 means the speed on this TMC segment in 35min advance is selected to predict the travel rate of the corridor of study.

The results are compelling. Those 18 speed features selected by LASSO can be categorized into the following groups:
\begin{itemize}
	\item Segments on I-270 northbound, upstream and downstream of the corridor.
	\item Segments on the alternative route (MD 335 North).
	\item Segments on I-495 North that merge into I-270 northbound.
	\item One segment on East West Highway with three different time lags is selected, and one segment of I-495 North Eastbound.
	\item Segments on several interchanges to the upstream of I-270 northbound.
\end{itemize}
Overall, the first three groups of features are expected by our feature selection criteria described in Section \ref{221}. The first group are fairly close to the targeted corridor in terms of degree of connections. Their correlations with the corridor can be explained by the propagation and spill back of congestion. For the two segments on MD 335 North in the second group, their correlations with the corridor are originated from the overwhelming travel demand from MD 335 to I-270 North. For the third group, since I-495 North is also the upstream of the corridor that serves one of the destinations of significant demand during afternoon peak, the traffic states on I-495 North can reveal the travel demand on I-270 in 30min advance. In addition, road segments in the fourth group are all eastbound. As their correlations with the northbound corridor are positive, we infer that the travel demand may peak at the same time for northbound and eastbound traffic during the afternoon peak. Last but not least, those segments on interchanges in the fifth group are all bottlenecks, as they are usually where congestion starts prior to congestion spills over to their upstream links. In other words, those segments on interchanges are more sensitive to incoming travel demand, and can serve as an early alert to the corridor of study. As a result, they are effective in predicting upcoming congestions.

\subsubsection{Prediction model}
\label{prediction}
We test and compare the performance of four prediction models: LASSO linear regression, stepwise regression, random forest and support vector regression. Model performance is evaluated by a 10-fold cross validation on each season (cluster). We adopt a univariate autoregressive moving average (ARMA) model as the baseline. It utilizes the speed data of the corridor only without considering spatio-temporal features of any kind. Prediction accuracy is measured by Normalized Root Mean Squared Error (NRMSE):
\begin{equation}
NRMSE  = \frac{\sqrt { \frac{1}{N_t}\sum_{t \in all} ({\hat{y}_t -y_t}})^2 }{\frac{1}{N_t}\sum_{t \in all} y_t}
\label{RMSE}
\end{equation}
In which $N_t$ is the total number of data points in a cluster, $\hat{y}_t$ and $y_t$ are the predicted and observed travel rate, respectively.

First, a random set of days are selected and used to find the best fit for the ARMA model based on the AIC values. The remaining days are used for a 10-fold cross validation to compute NRMSE. As a result, the best ARMA in this case is that the order of autoregressive and moving average is 3 and 3 respectively. When predicting travel time 5-min in advance, ARMA reaches 7.35\% in NRMSE. For 30-min ahead prediction, ARMA averages at 23.9\% in NRMSE. This shows that predicting travel time 30-min in advance is much more challenging than 5-min ahead.

The final selected features for prediction contain all 18 TMC features for each cluster, and all other features that are significant under their hypothesis tests in correlation analysis. Some common features for the two clusters are incidents on the downstream segments and on the alternative routes, morning and afternoon travel demand level, and essential weather features including visibility, precipitation type,  precipitation intensity, wind speed, and pavement conditions.

For stepwise regression, we use AIC values as the criteria. In LASSO, the norm-1 penalty $\lambda$ is set to maximize the percentage of deviance explained. The results of model fitting are shown in Table \ref{fitting}, in which CV training and CV test stand for the average training and testing errors in cross validation, respectively. Num.F stands for the average number of features used in each model, including the intercept (constant). Ave. CV test is the weighted average cross validation testing error of the two seasons, serving as an indicator for the final model performance. Results from random forest is based on the setting of 20 trees for each cluster. Note that by adjusting the number of trees in the model, its performance changes accordingly. In this case study, we test multiple values for the number of trees, ranging from 5 to 80, and find that as the number of trees increases from 5 to 20, the testing error improves from 17.8\% to 16.6\%, and levels off if the number of trees goes beyond 21.

By comparing the results of all models, we can see that the OLS regression on  multivariate speed features with clustering marginally improves the benchmark model, ARMA(3,3). Furthermore, by incorporating non-TMC based features, LASSO outperforms OLS regression while owning a much lower complexity. It is effective to incorporate non-TMC features in travel time prediction, and LASSO can ensure the prediction is more robust (less overfitting) than the OLS where a large number of features are used. In addition, the two linear regression models, LASSO and stepwise regression by AIC share similar performance in prediction with an average testing error of around 20.4\%. However, the number of finally selected features in the stepwise regression model is lower than LASSO, since it uses 31 and 29 features for two clusters, comparing to 47 used by LASSO. Finally, random forest outperforms other models considerably with an average error rate of 16.6\%. Comparing to the benchmark ARMA model, our model effectively reduces the prediction error by a margin of 7.2\%.

\begin{table*}[t!]
	\centering
	\renewcommand{\arraystretch}{1.2}
	\begin{tabular}{|l|l|l|l|l|l|l|l|}
		\hline
		\multirow{2}{*}{Model} & \multicolumn{3}{l|}{Winter} & \multicolumn{3}{l|}{Non-winter} & \multirow{2}{*}{Ave. CV test} \\ \cline{2-7}
		& Num.F  & CV train & CV test & Num.F   & CV train   & CV test  &                               \\ \hline
		Baseline--ARMA & \multicolumn{6}{l|}{NA} & 0.238 \\ \hline
		OLS on all TMCs        & 1231     & 0.162   & 0.230 & 1281       & 0.180   & 0.220        & 0.224 \\ \hline
		LASSO                   & 37       & 0.199   & 0.203 & 36         & 0.213   & 0.210        & 0.207 \\ \hline
		Stepwise AIC            & 31       & 0.196   & 0.196 & 29         & 0.208   & 0.210        & 0.204 \\ \hline
		Random forest           & 47       & 0.067   & 0.186 & 47         & 0.070   & 0.153        & 0.166 \\ \hline
		SVR                     & 47     & 0.160   & 0.182 & 47       & 0.169   & 0.182        & 0.182 \\ \hline
	\end{tabular}
	\caption{I-270 Case study: Model performance evaluations. Cross validation (CV) errors of predicting travel time 30-min in advance, errors are measured in NRMSE.}
	\label{fitting}
\end{table*}


\subsection{I-376 Eastbound}
The second case study is conducted on a road segment of I-376 Eastbound, between Forbes Ave exit and Squirrel hill exit in Pittsburgh Metropolitan Area. This 2.8-mile-long highway corridor is one of the main roadways connecting Pittsburgh downtown to the east region of the city. Due to heavy traffic load and limited roadway capacity, congestion is sensitive to demand and very frequent on this corridor during afternoon peaks. The time period of study is the same as the I-270 case study, 2PM-6PM on weekdays during the year of 2014, while feature selection is based on all the data in 2013. The stretch is marked in red in the left map of Fig. \ref{Pit_map}. We predict its travel time 30 minutes in advance.

\subsubsection{Data sources and feature engineering}
The following datasets were collected and used in this case study:
\begin{enumerate}
	\item Speed: TMC-based speed data from INRIX, in all 259 TMC segments covering all major roads near this corridor, in the neighborhood of Oakland and Southside. Historical data are available in 5-minutes granularity. Those TMCs are listed below and shown in the right map of Fig \ref{Pit_map}:
	\begin{itemize}
		\item Upstream/downstream of I-376 corridor of study, both Northbound and Southbound. (Blue on the map)
		\item Roadway network of the region, including Pittsburgh downtown, Oakland and Southside. (Red on the map)
		\item Three main arterial streets eastbound from Pittsburgh downtown:  Forbes Ave, Penn Ave and Center Ave, as alternative routes for I-376 during afternoon peaks. (Purple on map)
	\end{itemize}
	
	\begin{figure*}[t!]
		\centering
		\includegraphics[width=3.5 in]{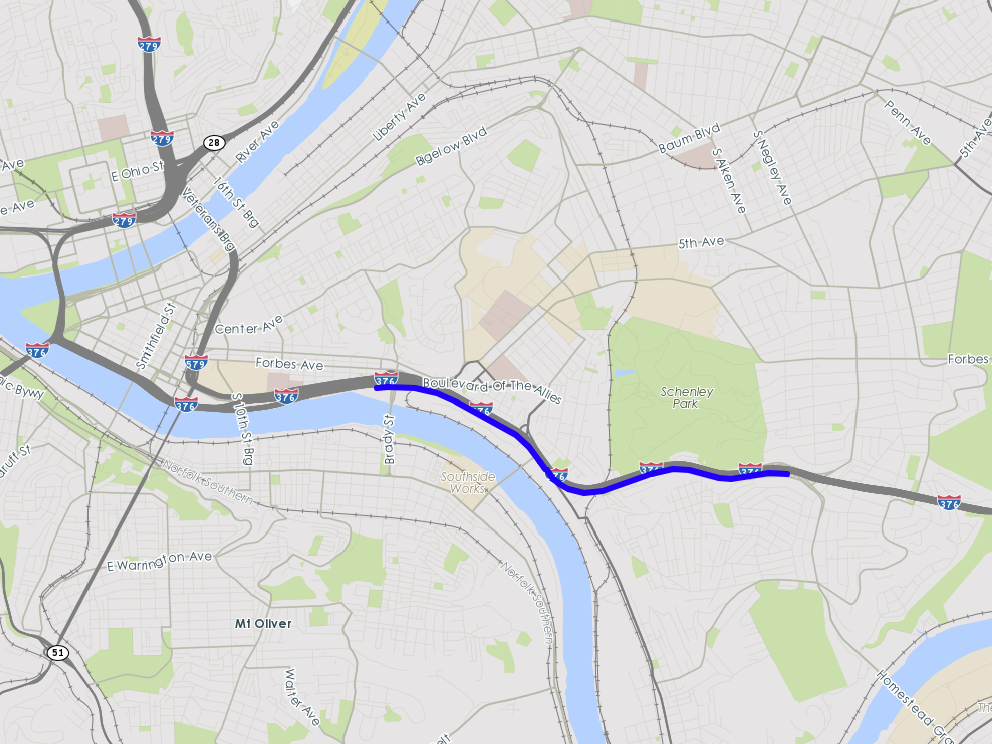}
		\includegraphics[width=3.5 in]{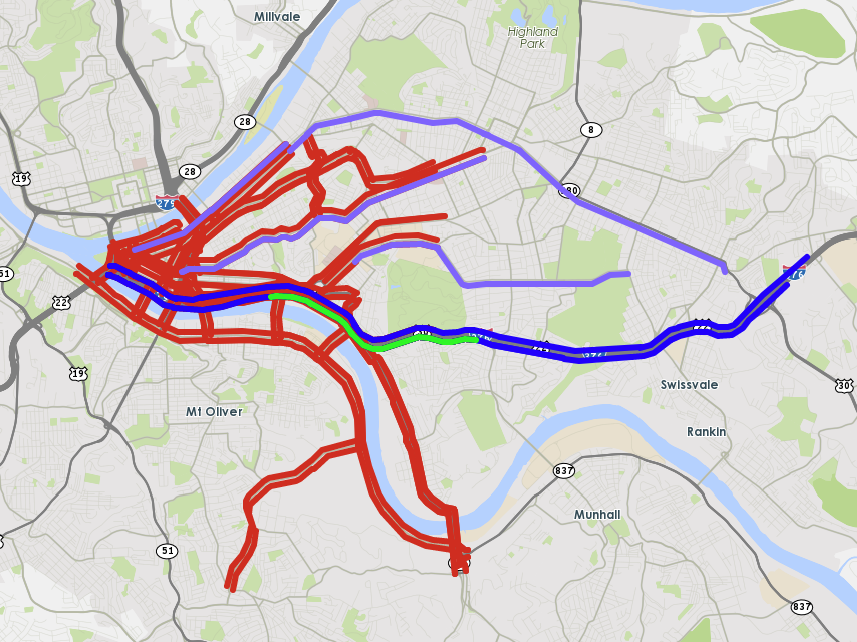}
		\caption{Left: I-376 Eastbound Segment of study; Right: Speed dataset used in the case study: 259 TMC segments in total.}
		\label{Pit_map}
	\end{figure*}	
	
	\item Incidents: incidents data are obtained from PennDOT Road Condition Reporting System (RCRS) where each incident entry is categorized into the following binary features based on its location:
	\begin{itemize}
		\item An incident on the upstream of the segments of study.
		\item An incident on the downstream of the segments of study, both severe and non-severe.
		\item An incident on the opposite direction (i.e., I-376 W).
		\item An incident on alternative routes (Penn Ave and Baum Blvd).
		\item An incident in Downtown Pittsburgh, far upstream of the segments of study.
	\end{itemize}		
	The definition of incident features in this study follows a similar fashion as in the I-270 case study, namely all features are binary and calculated based on the timestamp and geographic info of all RCRS entries. Due to the limited number of incident records on I-376, severe and non-severe incidents are aggregated into one feature in the model.
	\item Weather: the set of weather features is identical to the I-270 case study, including: temperature (degrees Fahrenheit), wind chill temperature, precipitation intensity (inch/hour), precipitation type (Snow/Rain/None), Visibility (miles), wind speed (miles/hour), wind gust (miles/hour), pressure (millibar), pavement condition type (wet or dry), as well as categorical features of wind, visibility and precipitation intensity.
	\item Local events: we incorporate the schedule of home games of NFL (Pittsburgh Steelers) and  NHL (Pittsburgh Penguins) to analyze their impacts. Similar to the I-270 case, the event feature is binary and constant for the entire PM peak, indicating whether there is an ongoing/incoming event on that day.
\end{enumerate}

\subsubsection{Clustering}
Clustering is conducted on a set of 35 TMC segments covering all major highways of the network, selected in a similar way to the I-270 case study. Speed measurements of three time points, 2:00PM, 4:00PM and 6:00PM are used to represent the traffic states of the network during afternoon peaks.

As a result of HAC , all weekdays of 2013 are split into the following two clusters:(1) 2013-01-01 to 2013-05-29 and 2013-12-10 to 2013-12-31; (2) 2013-05-30 to 2013-12-09. Unlike the case study of I-270, the separation of the two clusters can interpreted as winter/spring and summer/fall seasons respectively.

\subsubsection{Correlation analysis}
\begin{figure}[t!]
	\centering
	\includegraphics[width=5.5 in]{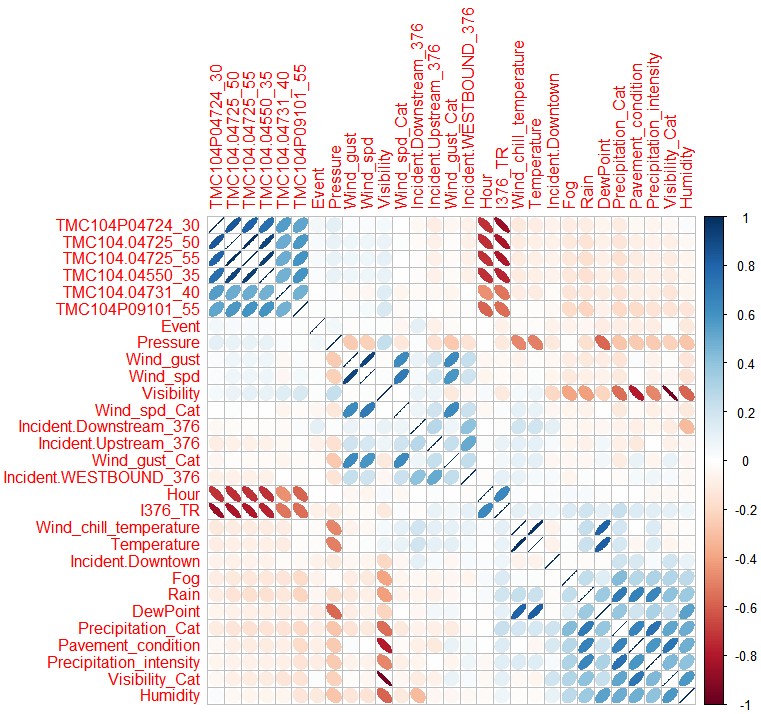}
	\caption{Correlation matrix of selected features for the winter/spring season.}
	\label{Corre_pit}
\end{figure}

We calculate the correlation matrix of the feature set for each cluster to explore the relationship among features, and conduct hypothesis tests to check whether certain features are linearly correlated. The correlation matrix for the winter/spring cluster is visualized in Fig. \ref{Corre_pit}. Similar to the I-270 case study, the correlations among top TMC features with the highest coefficients and the travel rate(travel time) of the corridor of study are significantly higher than other non-TMC features. Thu, those TMC speed data are the most important factors in explaining the variation of travel rate. Most non-TMC factors are correlated with the travel rate, such as incidents, visibility, drew points, pavement condition, the hour of day, and they should be included in the feature set for the prediction model. The features that are not significantly correlated with the travel rate (travel time) of the targeted corridor are thus removed from the feature set.

\subsubsection{Principal component analysis}
\begin{figure*}
	\centering
	\includegraphics[width=5.5 in]{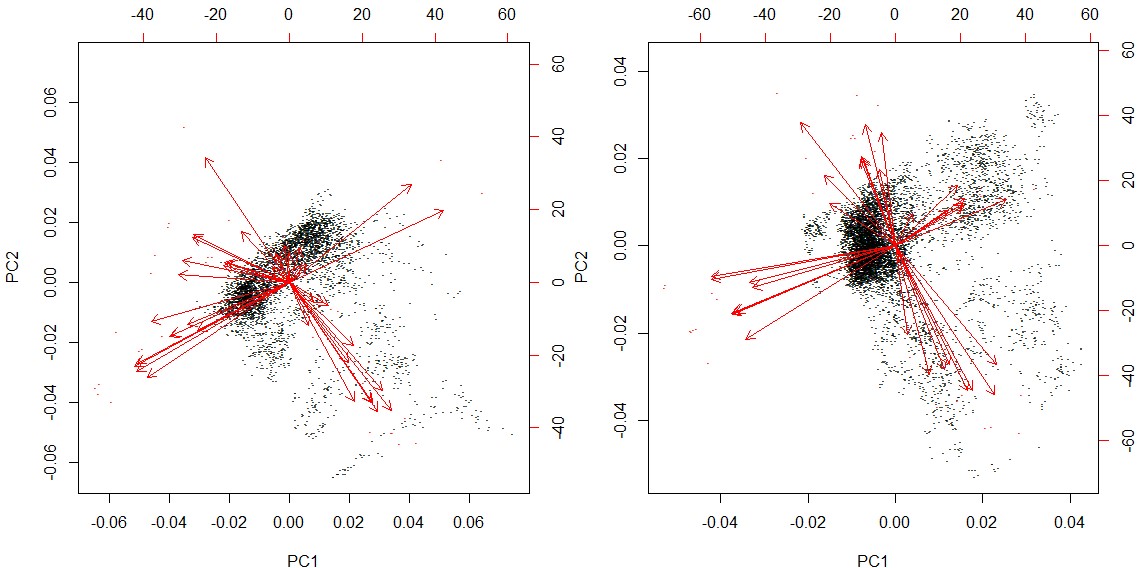}
	\caption{Plot of Principle components, Left: First half-year cluster; Right: Second half-year cluster. Each black dot stands for one data point mapped to the orthogonal space of the two PCs.}
	\label{PCA_pit}
\end{figure*}

Similar to the I-270 case study, PCA is conducted on the set of selected features for the two clusters, including the five most correlated TMCs, as well as features of traffic demand level, weather, incidents and events. The first two principal components (PC) are plotted in Fig \ref{PCA_pit}. The first PCs for both clusters contain several weather related features, including temperature, humidity, wind speed visibility and pavement condition, accounting for around 19\% of the total variance for each cluster. The second PCs consist of most TMC-based speed features, namely 6 TMCs for the first cluster and 9 for the second cluster, accounting for about 10\% of the total variance.

Comparing this result to the I-270 case study, it can be seen that the importance of non-TMC features differs from case to case, but the most correlated TMC-based speed features are always critical for prediction.

\subsubsection{Dimension reduction (feature selection) in the TMC-based speed data set}
We use LASSO to first select a subset of TMC-based speed features in predicting travel time/rate. A total of 15 TMC features for the first cluster and 17 for the second cluster are selected following a similar method discussed in Section \ref{select_DC}. These selected TMCs are then combined with other non-TMC features to create the feature set. LASSO is further used to select final features to be included in the prediction model.

\begin{figure}
	\centering
	\includegraphics[width= 5 in]{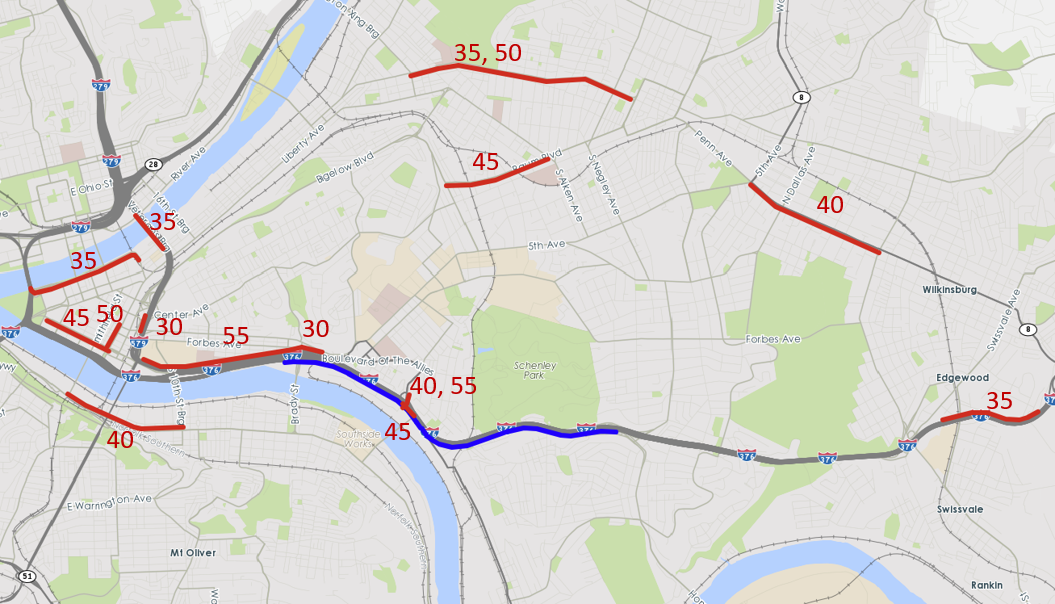}
	\caption{TMC segments selected for the winter/spring season. Time lags in minutes are listed for each selected TMC}
	\label{speed_PIt}
\end{figure}

In Fig \ref{speed_PIt}, the I-376 corridor of study are marked in blue, and the 15 selected TMC-based speed features for the first cluster are marked in red, along with time lags in minutes listed. Those 15 speed features can be categorized as follows:
\begin{itemize}
	\item A segment within the corridor of I-376 Eastbound.
	\item A segment on the ramp merging into I-376 Eastbound.
	\item Segments on the alternative routes: Penn Ave and Center Ave.
	\item Segments on the opposite direction (I-376 Westbound).
	\item Segments in Pittsburgh downtown area.
	\item A segment in one of the adjacent neighborhood, Southside.
\end{itemize}
The first two segments can be seen as direct indicators of overall congestion level for the entire corridor. They are likely two of the bottlenecks. The third group of segments being selected shows that high correlations from alternative routes can effectively help predict congestion on the targeted corridor. Again, this is proven to be useful for prediction similar to the other case study. The last two groups of segments imply that the corridor congestion may be related to those critical roadways from the neighborhoods that feed travel demand to it. Though this does not necessarily constitute causal relations, those segments can send signals 40min ahead to alert congestion on the corridor.
To sum up, the features selected for the final models of the I-376 case study include: 15 and 17 TMC features for the two clusters; essential weather features including visibility, precipitation type, precipitation intensity, wind speed, and pavement conditions; local events; incidents on up/down stream of I-376 E as well as incidents on I-376 W; Hour of day. 

\subsubsection{Prediction model}
To find the best prediction model for this case, we train and evaluate the following candidate models: ARMA as a baseline model, OLS regression on speed, LASSO, stepwise regression, random forest and support vector regression. The steps to run those models are the same as the I-270 case study, described in Section \ref{prediction}. For ARMA, the best fit in this case is two autoregressive terms and one moving average term. As the baseline, ARMA reaches an error rate of 14.22\% for 5-min ahead prediction and 38.4\% for 30-min ahead prediction.

All model results are compared in Table \ref{fitting_pit}. The ranking of all candidate models are quite similar to the I-270 case. The two linear regression models, LASSO and AIC-based stepwise regression share similar performance with an average testing error of 25.2\%, a significant improvement from ARMA, but not great yet. Random forest achieves the lowest average error again, at 17\% in NRMSE, with a 21\% improvement comparing to the baseline ARMA. By combining the results of the two case studies, we summarize that adding spatio-temporal information from all segments in the network, weather, incidents and events can greatly improve real-time prediction. On the other hand, we notice that the accuracy of each model is quite consistent in the two case studies. LASSO, SVR and random forest are all good methods, showing considerable improvement than ARMA and naive models that use speed data only.

\begin{table*}[t!]
	\centering
	\renewcommand{\arraystretch}{1.2}
	\begin{tabular}{|l|l|l|l|l|l|l|l|}
		\hline
		\multirow{2}{*}{Model} & \multicolumn{3}{l|}{Cluster 1(First half year)} & \multicolumn{3}{l|}{Cluster 2(Second half year)} & \multirow{2}{*}{Ave. CV test} \\ \cline{2-7}
		& Num.F  & CV train & CV test & Num.F   & CV train   & CV test  &                               \\ \hline
		Baseline--ARMA & \multicolumn{6}{l|}{NA} & 0.384 \\ \hline
		OLS on all TMCs   & 1063 & 0.219 & 0.270 & 1063 & 0.212 & 0.292 & 0.282 \\ \hline
		LASSO          & 31   & 0.247 & 0.260 & 33   & 0.233 & 0.245 & 0.252 \\ \hline
		Stepwise AIC   & 29   & 0.225 & 0.261 & 31   & 0.226 & 0.243 & 0.252 \\ \hline
		Random forest  & 36   & 0.080 & 0.164 & 38   & 0.082 & 0.175 & 0.170 \\ \hline
		SVR            & 36   & 0.162 & 0.193 & 38   & 0.190 & 0.208 & 0.200\\ \hline
	\end{tabular}
	
	\caption{I-376 Case study: Model performance evaluations. Cross validation (CV) errors of predicting travel time 30-min in advance, errors are measured in NRMSE.}
	\label{fitting_pit}
\end{table*}

\section{Conclusions}
We propose a data-driven method for analyzing highway congestion and predicting travel time based on spatio-temporal network characteristics and multiple data sources including travel speed, counts, incidents, weather and events, all in the context of dynamic networks. The proposed method can be used to analyze the spatio-temporal correlations among various features related to travel time, explore possible causal relations to congestion, and identify the most critical and reliable features for real-time travel time prediction.

The proposed method is applied to two regional highway corridors, I-270 in D.C. region and I-376 in Pittsburgh region. The results validate the effectiveness of the data-driven approach in understanding the correlations of highway congestion and various spatio-temporal features. We are able to predict travel time on those corridor 30min in advance, and the prediction results are satisfactory. In particular, we find that: 1) The days of year can be clustered into seasons, each of which show different traffic patterns; 2) TMC-based speed features are the most critical components of travel time variability in the multi-source data set. They include road segments on the alternative routes to the corridor of study, downstream and upstream bottleneck and major demand sources, all can be machine selected by the data-driven approach; 3) Other features that are useful in predicting travel time include time/location of incidents, morning and afternoon travel demand level, visibility, precipitation intensity, weather type(rain, snow), wind speed/gust, and pavement conditions; and 4) Random-forest shows the most promise of all candidate models, reaching an NRMSE of 16.6\% and 17.0\% respectively in afternoon peak hours for the entire year of 2014.


\section*{Acknowledgements}
This research is funded in part by Traffic 21 Institute and Carnegie Mellon University’s Mobility21, a National University Transportation Center for Mobility sponsored by the US Department of Transportation. The data acquisition and pre-processing for the I-270 corridor are funded by a FHWA research project ``Data Guide for Travel Time Reliability''. The authors wish to thank John Halkias, Douglas Laird, James Sturrock and David Hale for their valuable comments. The contents of this report reflect the views of the authors only. The U.S. Government assumes no liability for the contents or use thereof. 

\bibliographystyle{unsrt}
\bibliography{Reliability}
\end{document}